# Analysis of disfluencies for automatic detection of Mild Cognitive Impartment: a deep learning approach

K. López-de-Ipiña, U. Martinez-de-Lizarduy, P. M. Calvo, B. Beitia, J. García-Melero
Faculty of Engineering, Donostia-San Sebastian Spain
Miriam Ecay-Torres, Ainara Estanga
Fundación CITA Alzheimer, 20009 Donostia

Marcos Faundez-Zanuy
Escola Superior Politècnica de Mataró (UPF), Tecnocampus 08302 Mataró

*Abstract*—: The so-called Mild Cognitive Impairment (MCI) or cognitive loss appears in a previous stage before Alzheimer's Disease (AD), but it does not seem sufficiently severe to interfere in independent abilities of daily life, so it usually does not receive an appropriate diagnosis. Its detection is a challenging issue to be addressed by medical specialists. This work presents a novel proposal based on automatic analysis of speech and disfluencies aimed at supporting MCI diagnosis. The approach includes deep learning by means of Convolutional Neural Networks (CNN) and non-linear multifeature modelling. Moreover, to select the most relevant features non-parametric Mann-Whitney U-testt and Support Vector Machine Attribute (SVM) evaluation are used.

## I. Introduction

The World Alzheimer Report 2015 highlights that about 900 million people can be considered as the world's elderly population, and most of them live in developed countries [1]. AD is characterized by a progressive and irreversible cognitive deterioration including memory loss and impairments in judgment and language, along with other cognitive deficits and behavioral symptoms. An early and accurate diagnosis of AD helps patients and their families to plan the future, and offers the best possibilities of symptoms being treated. The cognitive loss appears in a previous stage, the so-called Mild Cognitive Impairment (MCI), but it does not seem sufficiently severe to interfere in independent abilities of daily life, so it usually does not receive an appropriate diagnosis. Its detection is a challenging issue to be addressed by medical specialists [2]. Along with memory loss, one of the major problems of AD is the loss of language skills. This loss is reflected in difficulties speaking to and understanding other people, which makes even more difficult the natural communication process and social interactions. This inability to communicate appears in early stages of the disease due to language difficulties, and it leads to social exclusion of people with AD, and to a serious negative impact not only on the sufferers, but also on their relatives [3]. In this sense, disfluencies are interesting language elements, which could be very useful to properly detect MCI. Both speech silences and disfluencies have valuable information for decoding the meaning of the uttered message.

The main aim of this project is the development of an automatic analysis of standard assessment tests such as categorical verbal fluency (CVF) by means of speech therapy techniques that will allow to obtain reliably and quickly these specific analyses [4]. Last years several works in the state of the art have addressed this. In the present paper we focus on the integration of language independent methodologies in order to detect AD in speech, with a task of Categorical Verbal Fluency (CVF), that is called animals naming task.

In section II the materials are described. The used methods are presented in section III. Section IV comprises the results and discussion, and finally in section V concluding remarks are drawn.

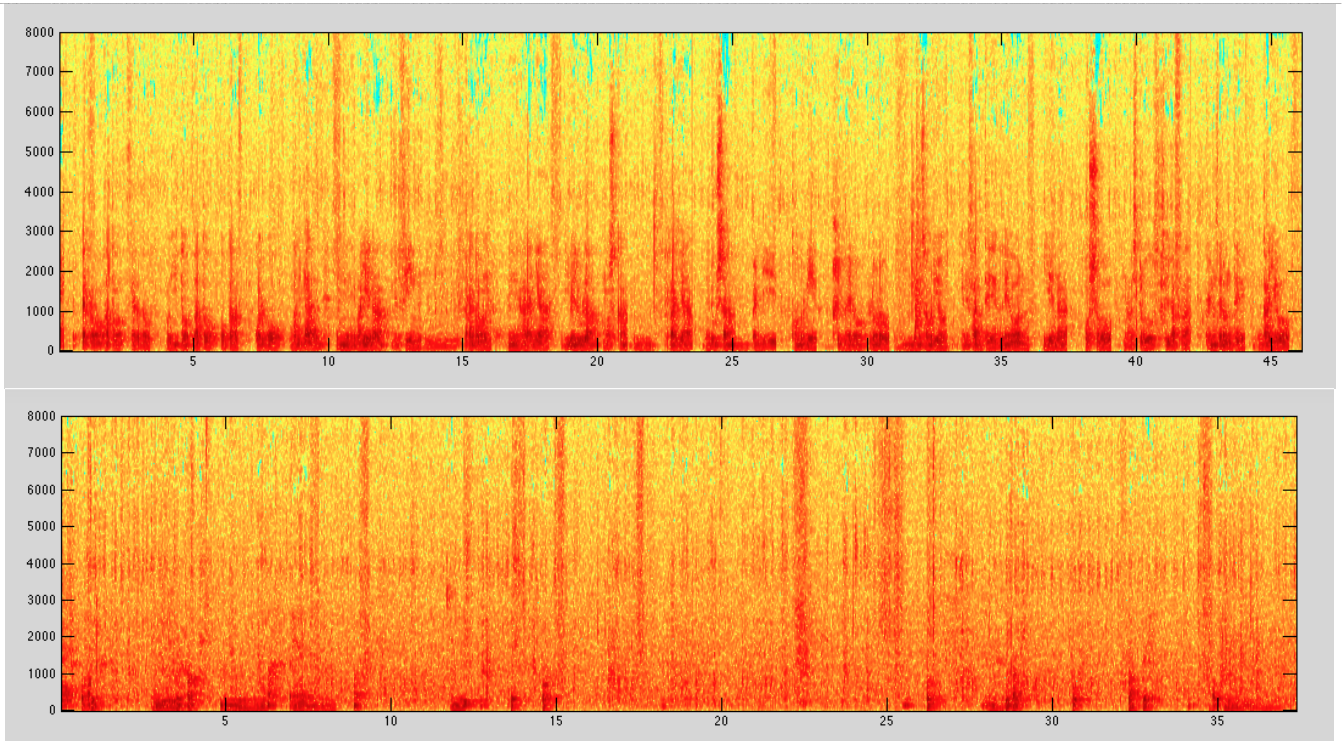

**Figure 1.** Categorical Verbal Fluency (CVF) task, for an individual of the MCI group (a) speech, (b) disfluencies

## II. MATERIALS

Recent works point out the relevance of disfluencies in speech to identify MCI and AD. In [5] it is suggested that shorter recording times reflect that AD patients require a greater effort to produce speech than healthy individuals: AD patients speak more slowly with longer pauses, and they spend more time finding the correct word, which leads to speech disfluencies or broken messages. Speech disfluencies are any break, irregularity or non-lexical element which occurs within the period of fluent speech, and that could start or interrupt it. These include among others: false starts, repeated or re-started phrases, repeated or extended syllables, grunts or non-lexical utterances such as fillers and repaired utterances, and instances of speakers correcting their own slips of the tongue or mispronunciations [6]. An increase in these disfluencies could be a clear sign of cognitive impairment, and in AD patients sometimes they become a verbal utterance of the internal cognitive process or an inner dialogue: "What is that?", "What was the name?", "/uhm/ "I can´t remember". The increase in the number of disfluencies and silences may point to a worsening of the disease, and could lead to a deficit in clear communication. As a conclusion, disfluencies are a direct reflection of the cognitive process during communication, and convey an unquestionable characteristic for the detection of these disorders. Although AD is mainly a cognitive disease, it may have phonation and articulation biomechanical alterations. The Categorical Verbal Fluency task (CVF, animal naming, AN), or animal fluency task, is a test for neurodegenerative diseases that measures and quantifies the progression of cognitive impairment [7], Fig. (1). It is commonly used in order to assess language skills, semantic memory and executive functions [8]. The sample includes 187 healthy people and 38 MCI patients that belong to the cohort of Gipuzkoa-Alzheimer Project (PGA) of the CITA-Alzheimer Foundation [4,9], Table (1). A balanced subset PGA-OREKA was selected for experimentation.

**Tab. (1).** Demographic data of the subsets selected for the experimentation

| | Female | Male | Range of age | Age-Mean | Age-SD |
|---|---|---|---|---|---|
| PGA-OREKA AN task-subset (CR/MCI) | 36/21 | 26/17 | 39-74/42-79 | 56.73/57.15 | 7.8/8.9 |

### III. METHODS

The new speech analysis approach is based on the integration of several kinds of features in order to model speech and disfluencies: linear and non-linear ones. Moreover, this approach is based on the description of speech pathologies with regard to phonotation, articulation, speech quality, human perception, and the complex dynamics of the system. In this work, we will use some of the most used speech features for both differentiation between healthy and pathological speech [4,5,10,11], and discrimination through human perception. Most of them are well known in the field of speech signal processing, and thus for each parameter a reference is provided where a deeper description and further information can be found. All features are calculated using software developed in our research group [4,5], MATLAB [12] and Praat [13].

*A. Automatic disfluency segmentation*

The recording has been automatically segmented in speech signal and disfluencies by means of a VAD (Voice activity detection) algorithm.

*B. Feature extraction*

Next, the analyzed features are described.

- *Classical features (CF):*

1. Spectral domain features: harmonic to noise ratio (HNR), noise to harmonic ratio (NHR), harmonicity, pitch, jitter, shimmer, APQ (Amplitude Perturbation Quotient); spectrum centroid and formants and its variants (mean, median, min, max, mode, std) [4-5].
2. Time domain features: Voiced/unvoiced segments, breaks, ZCR (Zero-Crossing Rate) [4-5] and its variants.
3. Energy, intensity, short time energy, and spectrum centroid [7-10, 12, 13].

- *Perceptual features (PF):*

1. Mel Frequency Cepstrum Coefficients (MFCC): Human ear behaves as some filters, it only concentrates on some components of frequency. These filters are not spaced uniformly on the frequency axis. There are more filters at low frequencies, and fewer filters at high frequencies. This kind of performance is simulated by means of Mel-Frequency analysis and particularly Mel Frequency Cepstrum Coefficients (MFCC) [4,5,14].
2. Coefficients that provide information related to voice quality, perception, adaptation or amplitude modulation: MSC (Modulation Spectra coefficients), PLP (Perceptual Linear Predictive coefficients), LPCC (Linear Predictive Cepstral coefficients), LPCT (Linear Predictive Cosine Transform coefficients) and ACW (Adaptive Component Weighted coefficients), ICC (Inferior Colliculus Coefficients). These features are sometimes extended by their 1st and 2nd order regression coefficients ($\Delta$ and $\Delta\Delta$ respectively) [10,14].

- *Non-linear features (NLF):* Fractal Dimension, Shannon Entropy and Multiscale Permutation Entropy have been calculated [4,5,14].

*C. Automatic selection of features by Mann-Whitney U-test*

In this step the best features are automatically selected with respect to common significance level. Therefore, the automatic selection of features is performed by Mann-Whitney U-test being p-value < 0.1 in order to get a bigger set for the second feature selection phase [14].

*D. Automatic selection of features by WEKA*

Then a new selection phase is carried in WEKA by the *SVMAttributeEval* algorithm. This provides a selection by analyzing the features group.

*E. Feature normalization by WEKA*

During the preprocessing of data, all the features will be normalized by WEKA.

*F. Automatic classification*

Four classifiers will be used: k-nearest neighbors (k-NN), Support Vector Machines, Multilayer Perceptron (MLP) with *L* layers and *N* neurons, and a Convolutional Neural Network (CNN) with *L* layers of *N* neurons, a convolution of *cxc* and a pool of *pxp*. We have used the WEKA software suite [15] to perform the experiments. Classification Error Rate (CER, %) has been used to evaluate the results. We have used k-fold cross-validation with k=10 for training and validation [15,16].

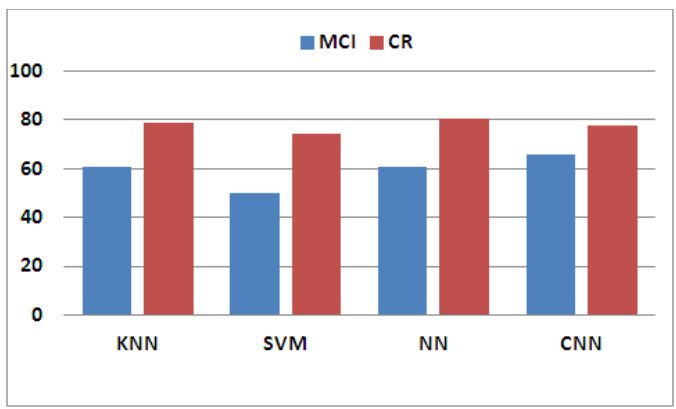

**Figure 2.** CER (%) for classes and selected classifiers: k-nearest neighbors (k-NN), Support Vector Machines (SVM), Multilayer Perceptron (NN) with *L* layers and *N* neurons and a Convolutional Neural Network (CNN)

## IV. RESULTS

In the experimentation the used materials are about 40 speech samples for the MCI group and 60 for the control group (CR), PGA-OREKA (Table 1). Initially, the obtained number of features is about 920 (473 for speech and 447 for disfluencies) for a sampling frequency of 22000 KHz. Then, after a normalization test, an automatic feature selection is performed based on a non-parametric Mann-Whitney U-test with a p-value < 0.1, and about 150 features are selected. In the second optimization step, the attribute selection algorithm *SVMAttributeEval* of WEKA yields about 80 features that are finally selected. The proposed feature set includes features from all the feature types described in subsection III.B for speech and disfluencies. Figure 2 shows the results of the automatic classification by classes, CR and MCI. CER (%) is evaluated for all the classifiers detailed in subsection III.F. The new proposal that includes integration of disfluency analysis outperforms previous works [4] for most of the classifiers. The results are hopeful, stable, good and equilibrated for all of them about %75. The deep learning option with CNN yields the best results for a configuration of 2 layers of 20 neurons, a convolution of 3x3 and a pool of 2x2. This option outperforms MLP for 2 layers of 100 neurons.

## V. CONCLUSIONS

This work presents a novel proposal based on automatic analysis of speech and disfluencies in order to support MCI diagnosis. A non-linear multifeature modeling is presented based on selection of the most relevant features by statistical tests (under medical criteria) and automatic attribute selection: Mann-Whitney U-testt and Support Vector Machine Attribute (SVM) evaluation. The approach includes deep learning by means of Convolutional Neural Networks (CNN). The results are hopeful and open a new research line.

## ACKNOWLEDGEMENT

This work has been supported by FEDER and MICINN, TEC2016-77791-C4-2-R and UPV/EHU-Basque Research Groups IT11156.